# The Impact of UCMAS Training Program on Potentiating Cognitive Capacity among 9-12 Year-Old Primary Schoolers in Shiraz

Ali-Mohammad Kamali[1,2,5], Fatemeh Shamsi[1,2,5], Zahra Zeraatpisheh[1,2,5], Mohammad-Mojtaba Kamelmanesh[3,4], Mohammad Nami[1,2,5,6*]

[1] Department of Neuroscience, School of Advanced Medical Sciences and Technologies, Shiraz University of Medical Sciences, Shiraz, Iran
[2] Neuroscience Laboratory (Brain, Cognition and Behavior), Department of Neuroscience, School of Advanced Medical Sciences and Technologies, Shiraz University of Medical Sciences, Shiraz, Iran
[3] Islamic Azad University, Fars Branch, Shiraz, Iran
[4] Pouya-Zehn-e-Bartar Educational and Scientific Institute, Shiraz, Iran
[5] Dana Brain Health Institute, Iranian Neuroscience Society, Fars Chapter, Shiraz, Iran
[6] Academy of Health, Senses Cultural, Sacramento, CA, USA

*Corresponding author:* Mohammad Nami MD, PhD, Department of Neuroscience, School of Advanced Medical Sciences and Technologies, Shiraz University of Medical Sciences, Shiraz, Iran. torabinami@sums.ac.ir

## Abstract

**Background and Objective:** The Universal Concept of Mental Arithmetic System (UCMAS) is a modern representation of an ancient art of mental arithmetic. Although such a training is hypothesized to result in potential effects on mental capacity and cognitive performance in experienced trainees, this has not sufficiently been systematically studied. The present study was an attempt to compare the objective testing score of Less-trained (≤6 months of training) and experienced (≥36 months of training) UCMAS-trained young adolescents in Shiraz.
**Materials and Methods:** Thirty healthy participants aged 9-12 were recruited from UCMAS training centers in Shiraz. The two study arms (comprising 15 Less-trained and 15 experienced children) were ensured to be sex- and age-matched. The Cambridge Brain Science-Cognitive Platform (CBS-CP) was employed as a media-rich computer testing battery for cognitive assessments. Moreover, cerebral blood levels of the participants was recorded through Hemoencephalography upon taking CBS-CP tasks (ANI task of each participant).
**Results:** The experienced UCMAS-trained participants were found to outperform in the Spatial Span (p=0.004), Digit Span (p=0.014) and Monkey Ladder (P=0.022) tests in which Short-Term Memory and Visuo-Constructive skills play important roles. Nevertheless, our findings revealed no statistically significant difference between-group differences in test scores which measured other cognitive domains.
**Conclusion:** Adding to the existing body of evidence on the effect of UCMAS training on potentiating cognitive capacity, our findings suggest that experienced UCMAS trainees tend to outperform in some key cognitive domains including Short-Term Memory and Visuo-Constructive Capacity. Further experimental neuroscience studies using quantitative electroencephalography and functional magnetic- of optic-neuroimaging would shed further light to neurodynamics of such a differentiation.

## *1. Introduction*

Performing precise numerical computation is a unique characteristics of adult humans which distinguishes them from other animals (1). Several devices such as paper and pencil, calculators and computers are used by most people for this purpose(2). Another device which is still widely used in Asian countries for arithmetic calculations is an abacus, a simple and traditional instrument composed of beads and rods in which the location of beads represents numbers (3).

Several arithmetic operations including addition, subtraction, division, root, etc can be done using an abacus by well-trained abacus users. Some such persons are so skillful that are able to perform calculations not only physically but also mentally using an imagery abacus. This skill is called abacus-based mental calculation(4) and is known as Universal Concept of Mental Arithmetic System or UCMAS nowadays. People with an extraordinary ability in performing both abacus-based mental and physical calculations because of long-term training are called abacus experts. In UCMAS, probably, these people have the ability of moving the imagery abacus beads in their mind faster than that of a physical abacus(5). In UCMAS training, mental arithmetic training such as working memory is included (6). This is a well-known idea that practice and experience can cause fundamental changes in the organization of cerebral cortex in adults. From the cognitive viewpoint, finding a mental solution for arithmetic problems requires integration of several cognitive functions including identification and manipulation of numbers in working memory, temporary storage and then retrieval meanwhile implementing basic arithmetic rules for control of different steps(7).

The neural mechanisms involved in mental calculation have been studied in UCMAS experts by some fMRI studies. An association between digit working memory and mental computation and an increase in neural resources for processing visuospatial information in adult experts of UCMAS has been reported by some studies. Furthermore, a study reported more fractional anisotropy in white matter of brain in children who were trained in UCMAS compared to untrained children using diffusion tensor imaging. According to previous studies, enhanced memory capacity and improved integrity of white matter tracts especially those involved in motor and visuospatial processes may result from long-term UCMAS training started from childhood(4).

Given the influence of UCMAS training on improvement of individual's skills and increasing accuracy and decreasing reaction time, Li et al.(2013) evaluated the influence of UCMAS training on digit working memory in children. Measuring the brain activity patterns using fMRI in both trained and untrained children during digit working memory performance showed more activity in some brain areas including the right posterior superior parietal lobule, superior occipital gyrus and supplementary motor area in trained children compared to controls(8). In addition, EEG and ERP studies have shown that numerical information processing in UCMAS trained children becomes faster and automatic because of the improved relationship between symbolic representation and numerical value(9). In addition to neuroimaging studies, the influence of UCMAS training on cognitive brain functions has been evaluated by some behavioral studies. In a study by Na et al. (2015), the effect of UCMAS training on memory, attention and arithmetic abilities was evaluated; better arithmetic ability and fewer commission errors were observed in UCMAS group compared to the control (10). Furthermore, evaluation of UCMAS training effect on math ability and executive function after 1 and 3 years of training showed that the performance of trained children in math and visuospatial domains was better than that of their untrained peers(11).

According to previous studies, it seems that UCMAS training can enhance cognitive functions by reinforcement of neural mechanisms underlying visuospatial attention. So, the present study was aimed to compare the objective testing score of Less-trained (≤6 months of training) and experienced (≥36 months of training) UCMAS trained children in Shiraz using Cambridge Brain Sciences Computerized Platform.

## 2. Materials and Methods

### 2.1. Subjects

30 participants aged 9-12 were recruited from UCMAS training centers in Shiraz and were classified into two groups according to their level of UCMAS training, experienced and Less-trained 15 children had received at least 36 months of UCMAS training and were assigned as experienced and 15 children had received less than 6 months of UCMAS training and were assigned as Less-trained. Prior to test, the study protocol was fully explained to the children and their parents and their entourage and written informed consents were obtained from all participants. This study was approved by The Ethics Committee of Shiraz University of Medical Sciences and it was conducted in accordance with the Declaration of Helsinki. All subjects were normal children without developmental, psychiatric and behavioral disorders.

### 2.2. Instrument

Cambridge brain science-Computerized Platform (CBS-CP) battery was used to assess different mental capacities including memory, problem solving, reasoning, and verbal ability. CBS is a computerized test including twelve tasks assessing three components of reasoning, memory and verbal abilities. In this study, two tasks of verbal ability were excluded because they were in English and the participants were not familiar with English language.

Before each task, the task was clearly explained to the participants by the examiner and they were asked to perform it once under the supervision of the examiner as the training trial. After making sure that there was no misunderstanding, the subject was asked to perform the main trial. This procedure was repeated for all ten tests. The tests were performed in a quiet room with sufficient illumination. After that, the participants' performance on the tasks were evaluated and the area needing improvement (ANI) was detected for each participant. The participants were asked to perform the ANI task again, this time the cerebral blood flow of the patiicpants in FP1 was recorded.

The following tests were administered to assess the several brain cognitive functions of the participants. Odd one out and rotation are to assess reasoning, polygon and feature match are designed to assess concentration, spatial planning to assess planning, spatial span and paired associations, digit span, token search, and monkey ladder are supposed to evaluate working memory (12) .

Spatial span is a test that mostly relies on the participant's short term memory capacity especially visuospatial working memory. In this test a sequence of flashing boxes appear on the screen and the participant must remember and click on the boxes on the same order that they had been flashed. If the participant is correct, one box will be added to the next problem. Accuracy but not speed is important for getting maximum score in this test. (CBS)

Monkey ladder is a test mostly dependent on the participant's short tern memory. Some boxes are presented on the screen at random location each containing a number. After a short time the numbers will disappear and the participant must remember the location of the numbers and click the boxes in an ascending numerical sequence starting from the box containing 1.

Digit span is a test that mainly relies on the participant's verbal ability and to lower extent on short-term memory. A number of digits are appeared on the screen one at a time and the participants must memorize them in the other they have been shown and after disappearing the last number, he/she must click on the number have been shown on the same order.

In Odd one out, the subject is required to choose the figure which is different from others in terms of shape, color or the number of copies according to a rule. A 3 × 3 grid of cells is displayed on the screen showing a variable number of copies of a colored shape in each cell. The objects in each cell are related to each other according to a set of rules(13). To find the different object the participant must deduce the rules that relate the objects. The test starts with simple rules but its difficulty increases with every correct answer 39. For more difficult problems participants are required to pay attention to more than one feature at a time.

Token search is a test for assessing working memory as well as visual search. The subject is required to memorize the sequence of signs and memorize their locations40. In this test some boxes appear on the screen, the participant must search boxes for a hidden token in one of them. When it was found the participant must remember its location and search for tokens in other boxes until one has been found in every box.

In paired association test some boxes are displayed on the screen and open one after another to reveal an enclosed object. After disclosing all objects, the objects are shown randomly in the center of the screen and the participant must click on the boxes which contained them. By completion of each trial correctly, the difficulty level of test increases by adding one object –box pair in the following trial(13). Accuracy is important in this test but speed does not matter. Short term memory has a great role in CBS-CP.

In feature match test, two sets of abstract shapes are appeared in two adjacent boxes on the screen. The participant must indicate if the contents of the two boxes are identical. If the participant is correct the complexity level of test increases in the next problem(13). Both accuracy and speed are important for getting the maximum score. This test is a perceptual test which requires the participants to concentrate and focus attention on complex images.

Spatial planning is a test which is based on the tower of London task and it is commonly used for measuring executive function. A number of beads are positioned on a tree-shaped frame and the participant is required to reposition the beads in a way to configure them in ascending numerical order from left to right and up to bottom on the frame(13). Both accuracy and speed are important and participants must try to solve the problems with the minimum possible moves, as fast as possible.

In rotation test two boxes containing red and green squares appear on the screen and the participant is required to judge if one of the boxes rotates would it be identical to the other one. This test is based on the participant's mental ability to rotate objects and it measures reasoning

Polygons is another test that mostly measures the participant's reasoning ability. Two panels appear on the screen that one contains two overlapping shapes and the other contains just one shape. The participant must identify if the single shape is identical to one of the shapes in the other panel.

### 2.3. Hemoencephalography

Hemoencephalography (HEG), a metabolic activity detection neuroimaging technique (Peanut nIR HEG kit) was used in this study to check cerebral blood flow in prefrontal cortex. The HEG data was recorded from both the Less-trained and experienced UCMAS trainees while preforming on area needing improvement (ANI) trials of CBS-CP.

### 2.4. Statistical analysis

The SPSS statistical package (Version 22.0.0, Copyright© IBM, 2015) 22 was used for data analyses. Normal distribution of data was tested by Shapiro-Wilk test. For normally distributed data, Independent sample t-test was used to compare means between two groups. For data that were not distributed normally, nonparametric tests were used.

## 3. Results

Significant differences were observed between the experienced and Less-trained in the following tests: spatial span (P=0.004), digit span (P=0.014) and Monkey ladder (P=0.022) in which short term memory plays an important role. Digit span test also relies on participant's

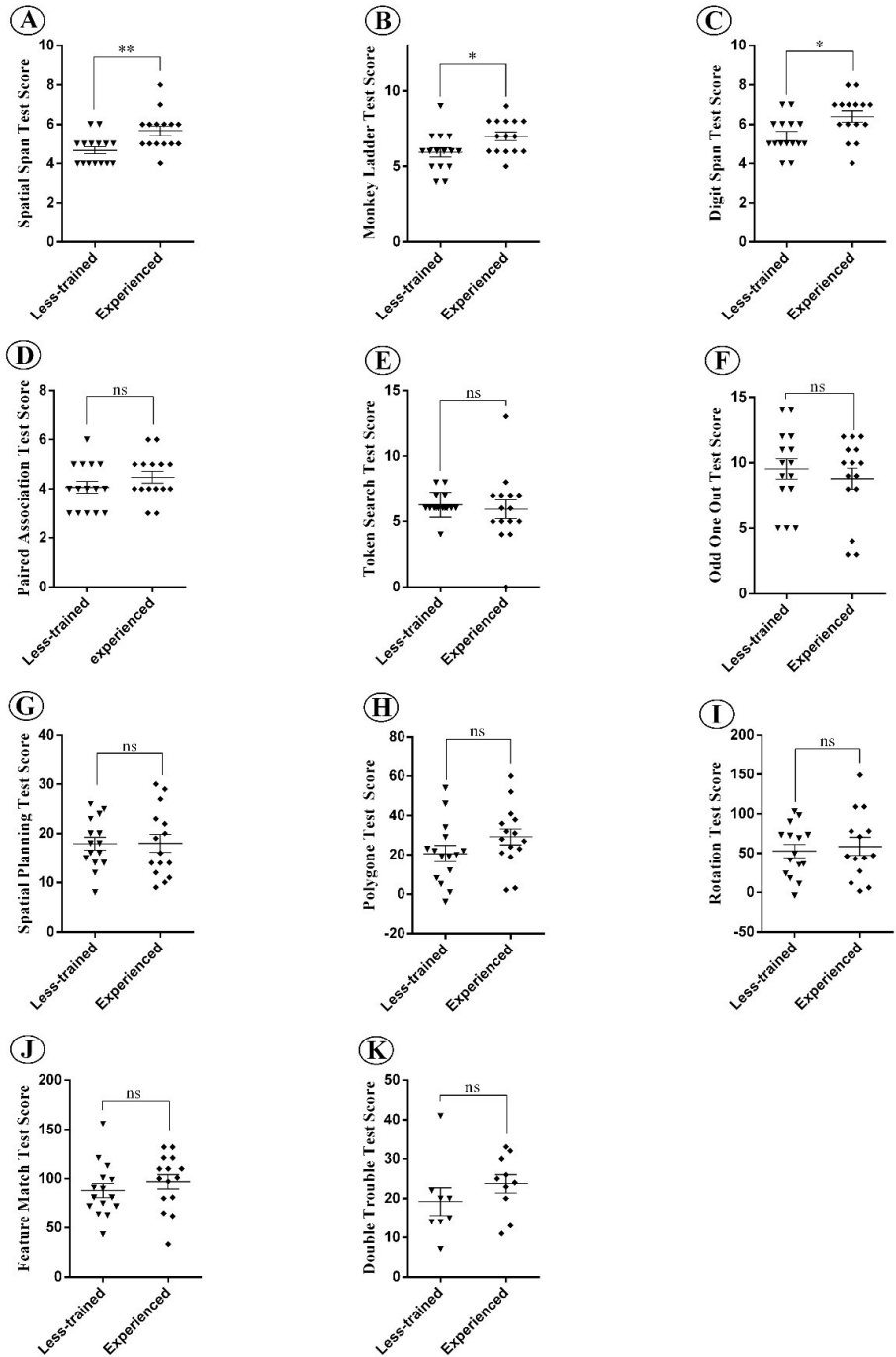

Figure 1. Scatter plots represent the performance of less-trained and experienced UCMAS trainees on CBS-CP trials. Panel a shows a significant difference between the performance of Less-trained and experienced subjects on Digit span (verbal task) (p <0.05). Panel b indicates a significant difference between Less-trained and experienced subjects upon visuospatial working memory task (Monkey ladder) (p<0.05). Panel c represents a significant difference in a memory task (Spatial span) scores from Less-trained to experienced trainees (p<0.05). Panel d indicates no significant difference between Less-trained and experienced trainees upon a reasoning task (Odd one out) (p<0.05). Panel e shows no significant difference between the performance of Less-trained and experienced subjects on Feature match (p<0.05). Panel f shows no significant difference between the performance of Less-trained and experienced subjects upon a memory task (Paired association) (p<0.05). Panel g reveals no significant difference between the performance of Less-trained and experienced subjects upon a self-ordered search task (Token search) (p<0.05). Panel h indicates no significant difference between Less-trained and experienced trainees upon a reasoning task (Rotation) (p<0.05). Panel i represents no significant difference between Less-trained and experienced trainees upon an executive function task (spatial planning) (p<0.05). Panel j shows no significant difference between the performance of Less-trained and experienced subjects on Polygons (p<0.05). Paired sample t-test was used with the $p$ value at .05. n. s., nonsignifican

verbal ability. There was no significant differences between the two groups in tests which measure reasoning including Odd one out (P= 0.601), Polygons (P=0.151) and Rotation (P=0.67) tests. Executive function measured by spatial planning also wasn't significantly different between the two groups (P=0.976). Also, no significant difference was found between the two groups in Feature match test which mostly assesses participant's concentration capacity (P= 0.39), token search results that mostly measures working memory and visual search were not significantly different between the experienced and Less-trained as well (P= 0.425), paired association test for which short term memory is important also showed no significant difference between the two groups (P=0.296) (Figure 1). In order to evaluate the effects of UCMAS training on cerebral blood flow, A paired-sample t-test was used to compare Δ optical density in FP1 between the performance of Less-trained and experienced subjects upon taking CBS-CP tasks (ANI task). The findings indicated a statistically significant increase in blood flow levels upon performance on ANI CBS-CP tasks in the experienced group ($p<0.05$) (Figure 2).

## 4. Discussion

The purpose of the present study was to compare the cognitive capacity of Less-trained and experienced UCMAS trained children. Mathematical training in UCMAS is based on a mental abacus (MA) the trainees keep in their mind while performing mathematical analysis. CBS trials applied in this study revealed that children under at least three years UCMAS training outperformed the control group in terms of Monkey ladder, Spatial span, and digit span tasks, the trials associated with working memory. Monkey ladder is mostly related to visuospatial working memory (15). Similarly, Spatial span is a task that mainly relies on the participants' short term memory, specifically visuospatial working memory. Being included in the verbal module, Digit span addresses short term verbal memory (16). Recruiting brain regions including visuospatial working memory, motor procedures, verbal processing and verbal working memory, MA training can probably enhance arithmetic abilities in trainees through focusing on visuospatial working memory circuits including hippocampal-prefrontal associations (17).

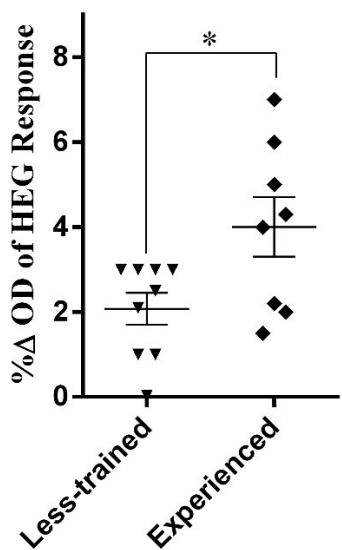

**Figure2.** Scatter plots show the performance of less-trained and experienced UCMAS trainees upon CBS-CP tasks recorded through HEG. Panel a indicates a significant difference between the percentage of ΔOD (optical density) of HEG response in the experienced versus Less-trained UCMAS trainees (p<0.05). Paired sample t-test was used with the p value at .05. n. s., non-significant

Two theories which can be discussed regarding this arithmetic enhancement are cognitive transfer leading to MA expertise through visual working memory manipulation and cognitive moderation relying on preexisting robust cognitive abilities, especially short-term working memory as moderators of efficient learning (Baron & Kenny, 1986). Confirming these two theories, it was shown that UCMAS training has significantly improved visuospatial and verbal working memory. Moreover, HEG results proved the effectiveness of UCMAS training on increasing FP1 optical density, presumably due to fronto-ocipital interactions during visual functioning. Ventro-medial prefrontal cortex and its connection to visual cortex may have mediated visual mechanisms involved in visual search and discrimination (18).

## *5. Concluding Remark*

Based on the obtained results from increased visual attention optical density in experienced trainees, working memory's close association with attention can be verified in agreement with the cognitive transfer hypothesis. MA expertise may rely on cognitive abilities such as manipulating imaginary beads, working memory, and attention for repeated practice of the procedures which are crucial for MA projection and further efficacy in MA learning and mastery.